\documentclass[prl,twocolumn,amsmath,amssymb,floatfix,reprint]{revtex4-1}

\usepackage{graphicx}
\usepackage{dcolumn}
\usepackage{bm}
\usepackage{color}
\usepackage[usenames,dvipsnames]{xcolor}
\usepackage{mciteplus}
\usepackage{braket}
\usepackage{ulem}

\def\AD#1{{\textcolor{OliveGreen}{#1}}}  

\begin{document}
\title{Preferential concentration of free-falling heavy particles in turbulence}
\author{F.~Falkinhoff$^{1,3}$, M.~Obligado$^2$, M.~Bourgoin$^3$, and P.D.~Mininni$^1$}
\affiliation{$^1$Universidad de Buenos Aires, Facultad de Ciencias Exactas y Naturales, Departamento de F\'\i sica, \& IFIBA, CONICET, Ciudad Universitaria, Buenos Aires 1428, Argentina.}
\affiliation{$^2$ Universit\'{e} Grenoble Alpes, CNRS, Grenoble-INP, LEGI, F-38000, Grenoble, France}
\affiliation{$^3$ Univ Lyon, Ens de Lyon, Univ Claude Bernard, CNRS, Laboratoire de Physique, 46 all \'ee d’Italie F-69342 Lyon, France}
\date{\today}

\begin{abstract}
We present a sweep-stick mechanism for heavy particles transported by a turbulent flow under the action of gravity. Direct numerical simulations show that these particles preferentially explore regions of the flow with close to zero Lagrangian acceleration. However, the actual Lagrangian acceleration of the fluid elements where particles accumulate is not zero, and has a dependence on the Stokes number, the gravity acceleration, and the settling velocity of the particles.
\end{abstract}
\maketitle

In spite of its apparent simplicity, the problem of spherical particles settling in a fluid hides a whole hierarchy of rich intricate phenomena, some of which are still shrouded in mystery, and which impacts numerous real situations. Atmospheric pollutants, cloud droplets, dust in proto-planetary accretion disks, sprays in engines are just examples pertaining to the broad class of turbulent flows laden with particles, which occur in many industrial and natural systems \cite{toschi2009lagrangian, balachandar2010turbulent}. In the context of pandemics, a current example of the importance of particle transport is given by the role of pollutants and aerosols as vectors of transmission and long range propagation of viruses \cite{newengjmed}. In these situations, the flow carrying the particles is turbulent. Unveiling the fundamental mechanisms driving the dynamics of transport and settling is therefore a crucial issue to improve our capacity to model and predict particle dispersion and deposition.

As surprising as this may seem, our modelling capacity is so weak that we are still unable to give quantitative answers to questions as simple as: Do small (point-like) spherical particles in a turbulent environment settle slower, faster, or at the same speed as in a quiescent fluid? What is, statistically, the spatial distribution of particles in a turbulent flow? And how is the distribution modified by gravity? The reasons are certainly related to the complexity of turbulence, one of the best known examples of out-of-equilibrium statistical systems, and to the difficulty added to the problem when the multi-scale and random dynamics of the flow is coupled to the particles' dynamics \cite{bourgoin2014focus, falkovich2002acceleration}. One of the most striking examples of this complexity is given by the phenomenon of preferential concentration: whereas turbulence is generally considered as a mixing enhancer, inertial particles in turbulence tend on the contrary to get \textit{unmixed}, and to concentrate in certain regions forming clusters. This effect impacts a whole range of the particles' dynamical features, as it changes their effective mean free path, impacting on cloud formation, particle aggregation, phase transitions, and predictions of local hazard thresholds 
\cite{monchaux2012analyzing, gustavsson2014clustering}.

In this letter we address the question of the mechanism driving preferential concentration of inertial particles in turbulence focusing on the interplay between clustering and settling. It has been observed in direct numerical simulations (DNSs) and experiments that preferential concentration is stronger when the Stokes number St (the ratio of the particle relaxation time to the Kolmogorov time) is close to unity \cite{Coleman2009, bec2007heavy}. The reason why turbulence affects the spatial distribution of particles is not completely clear, although an explanation is based on the centrifugal expulsion of heavy particles from turbulent eddies, that would result in the accumulation of particles in low-vorticity regions of the carrier flow \cite{Squires1994}. A more recent scenario, the so-called sweep-stick mechanism, was proposed in which particles cluster instead in regions of null Lagrangian acceleration \cite{goto2008sweep}. There has been growing evidence that for particles with $\textrm{St}<1$ there is a prevalence of centrifugal effects and particles cluster in low vorticity regions, whereas for $\textrm{St}>1$ the sweep-stick mechanism is more prominent and particles cluster in low Lagrangian acceleration points \cite{Coleman2009, Obligado2014}.

Nevertheless, these accumulation mechanisms do not take into account the effect of gravity, which is important when particles are heavy and can settle or precipitate. Gustavsson and coworkers \cite{gustavsson2014clustering} have shown that clustering properties may be significantly affected by gravity, pointing to the strong link between preferential concentration and settling. In the same context, an extension of the sweep-stick mechanism has been proposed, suggesting that settling particles concentrate in regions where the Lagrangian acceleration of the carrier flow equals that of the gravity \cite{hascoet2007turbulent}, although to our knowledge no experimental or numerical studies have explored this mechanism yet. More generally, the settling of inertial particles has been studied focusing on the possible enhancement or hindering by turbulence of the particles' terminal velocity (see, e.g., \cite{bib:good2014_JFM,bib:rosa2016_IJMF} for recent results), but except for a few recent studies \cite{bib:dejoan2013_PoF, Bec2014, sozza2016large, baker2017coherent, bib:monchaux2017_PRF, Saito2018} the interplay between preferential concentration and gravity for inertial particles has generally been neglected, either because it was not considered in simulations, or because it was negligible in the range of parameters considered in experiments. This is the case we focus on in our study. By studying the dynamics of particles in homogeneous and isotropic turbulence (HIT), we characterize preferential concentration, and pay particular attention to the values of the Lagrangian acceleration at the points where particles accumulate. We neglect situations in which particles and the fluid become two-way coupled, which may lead to subtle collective effects as experimentally observed in \cite{bib:aliseda2002_JFM}, and consider instead one-way coupling regimes in which the flow transports the particles without being perturbed by them, and in which accumulation results from modifications to the preferential sampling induced by gravity.

\begin{figure}
\includegraphics[width=8.8cm,trim={0 1.3cm 0 1.3cm},clip]{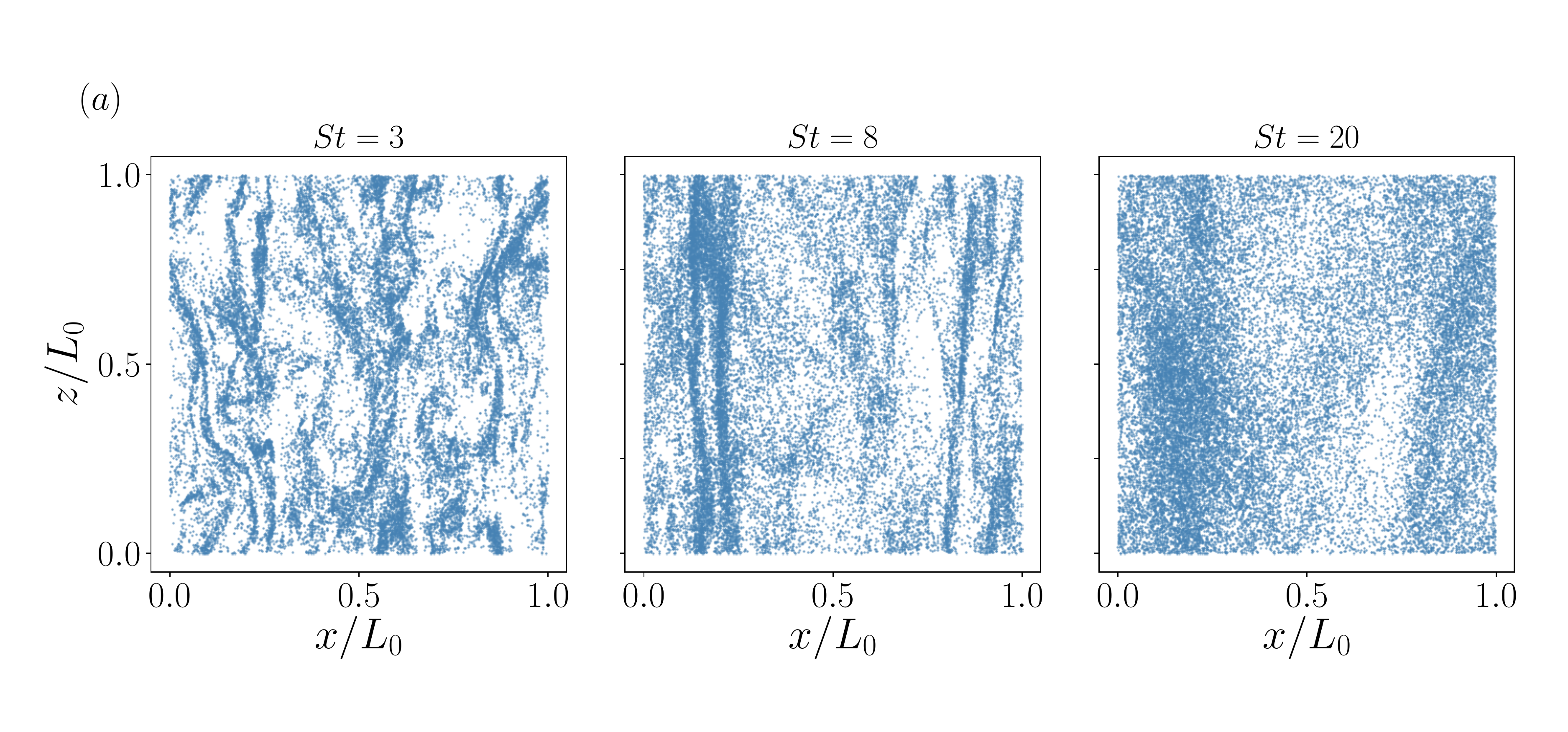}
\includegraphics[width=8.8cm,trim={.5cm 1.5cm .4cm .3cm},clip]{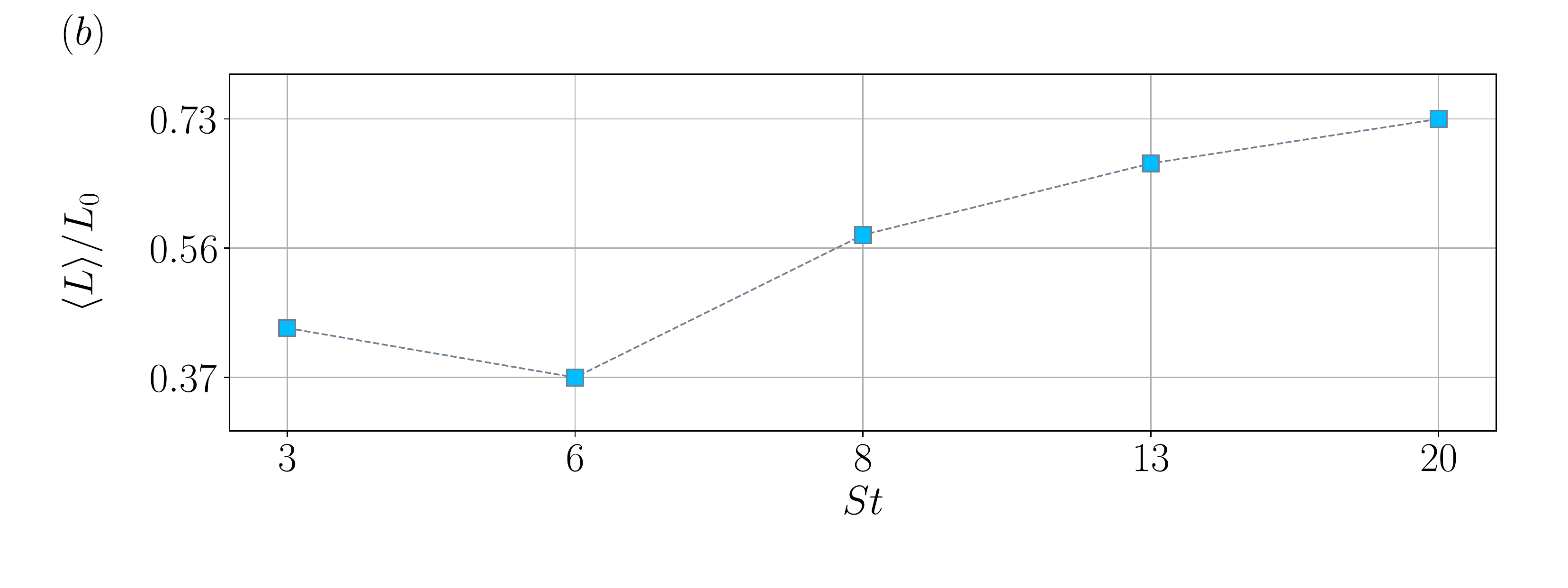}
\caption{{\it (a)} Particle distribution in an arbitrary $xz$ plane in simulations with $\textrm{Fr}=0.23$ and for different Stokes numbers $\textrm{St}=3$, $8$, and $20$. Columns form in the direction parallel to gravity. {\it (b)} Column width as a function of St, for $\textrm{Fr}=0.23$.}
\label{fig:columns}
\end{figure}

Using a pseudo-spectral code \cite{mininni2011hybrid} the velocity field in a three-dimensional periodic domain of length $L_0$ is obtained from DNSs of HIT, by solving the incompressible Navier-Stokes equation with $512^3$ grid points and an external mechanical forcing with random phases. The fluid viscosity $\nu$ is chosen so that the Kolmogorov scale is well resolved. The Taylor-based Reynolds number is $\textrm{Re}_\lambda \approx 300$; in the following, $U$ is a r.m.s.~flow velocity and $L$ the flow integral scale \cite{SM}. Inertial particles are modelled using the Maxey-Riley-Gatignol equation in the limit of point-heavy particles, which for a particle with velocity $\mathbf{v}$ at position $\mathbf{x_p}$ in a flow with velocity $\mathbf{u}$ is 
\begin{equation}
 \dot{\mathbf{v}}(\mathbf{x_p},t) = [\mathbf{u}(\mathbf{x_p},t)-\mathbf{v}(\mathbf{x_p},t)]/\tau - 
 g\hat{z},
 \label{eq:inertial}
\end{equation}
where $\tau$ is the particle Stokes time and $g$ an effective gravity that depends on the ratio between the densities of the particle and the carrier flow. For heavy particles, as the ones considered here, $g$ is approximately the gravity acceleration. With this equation particles in a fluid at rest reach a terminal velocity $v^* = -g\tau$ in a Stokes time.

The system has two dimensionless numbers: the Stokes number $\textrm{St}  = \tau/\tau_\eta$ that compares the particle relaxation time to the Kolmogorov time, and the Froude number $\textrm{Fr} = a_\eta/g = \varepsilon^{3/4}/(g\nu^{1/4})$ (with $\varepsilon$ the energy injection rate) that compares the turbulent acceleration at the Kolmogorov scale $a_\eta$ to the gravitational acceleration.  We study the statistics of inertial particles by injecting multiple sets of $10^6$ particles each into the turbulent flow. Between each set, $\tau$ and $g$ are varied to consider particles with different $\textrm{St}$ and $\textrm{Fr}$, resulting in 16 sets of particles, in all cases integrated for sufficiently long times for the particles to reach their terminal velocities and a statistical steady state. With the goal of exploring the validity of the sweep-stick mechanism in the presence of  gravitational forces, only particles with $\textrm{St}\geq 1$ are considered.

In the presence of gravity, particles precipitate at a mean velocity $\left<v_z\right>$ but still accumulate in preferential regions. While for $g=0$ clusters are isotropic, for $g$ sufficiently large particles fall through channels as seen in Fig.~\ref{fig:columns}(a) (also seen in \cite{gustavsson2014clustering}), facilitating concentration and cluster formation. For fixed Fr, the width of the channels depends on St, and for large St channels become wider with a
characteristic width close to the flow integral scale $L\approx L_0$ as shown in Fig.~\ref{fig:columns}(b) (indeed, for strong drag we observe that the distribution of particles becomes more homogeneous, where the characteristic scale of the channels $\langle L\rangle$ is defined as a mean correlation length for the vertically averaged particle density). Scaling laws for these structures were studied in detail in \cite{Bec2014}, but what is the mechanism behind their formation? Do particles sample regions of the flow preferentially? And does such sampling have an effect on particle settling velocities?

\begin{figure}
\includegraphics[width=8.6cm,trim={0 .1cm 0 .9cm},clip]{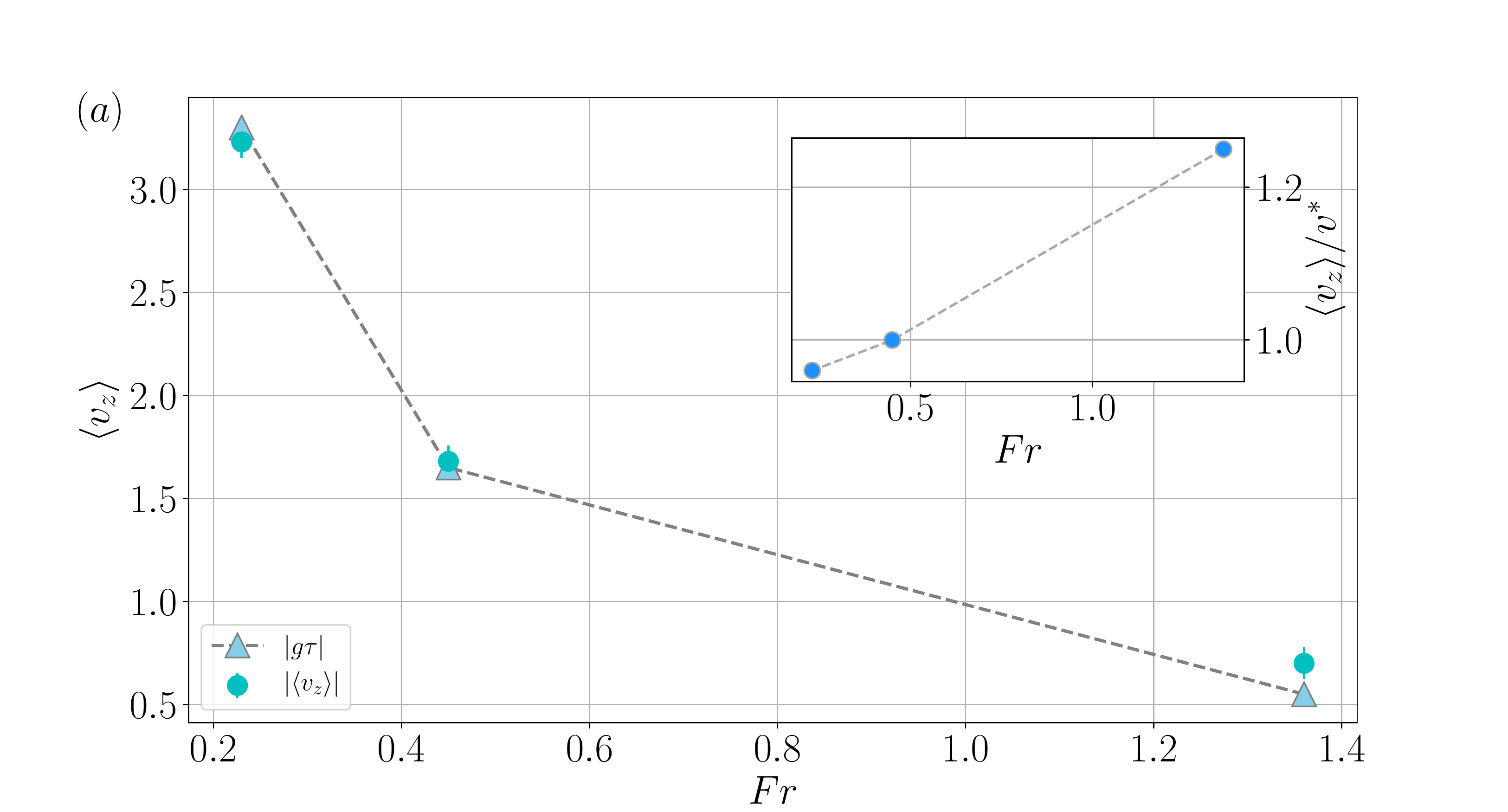}
\includegraphics[width=8.8cm,trim={0 .1cm 0 1.6cm},clip]{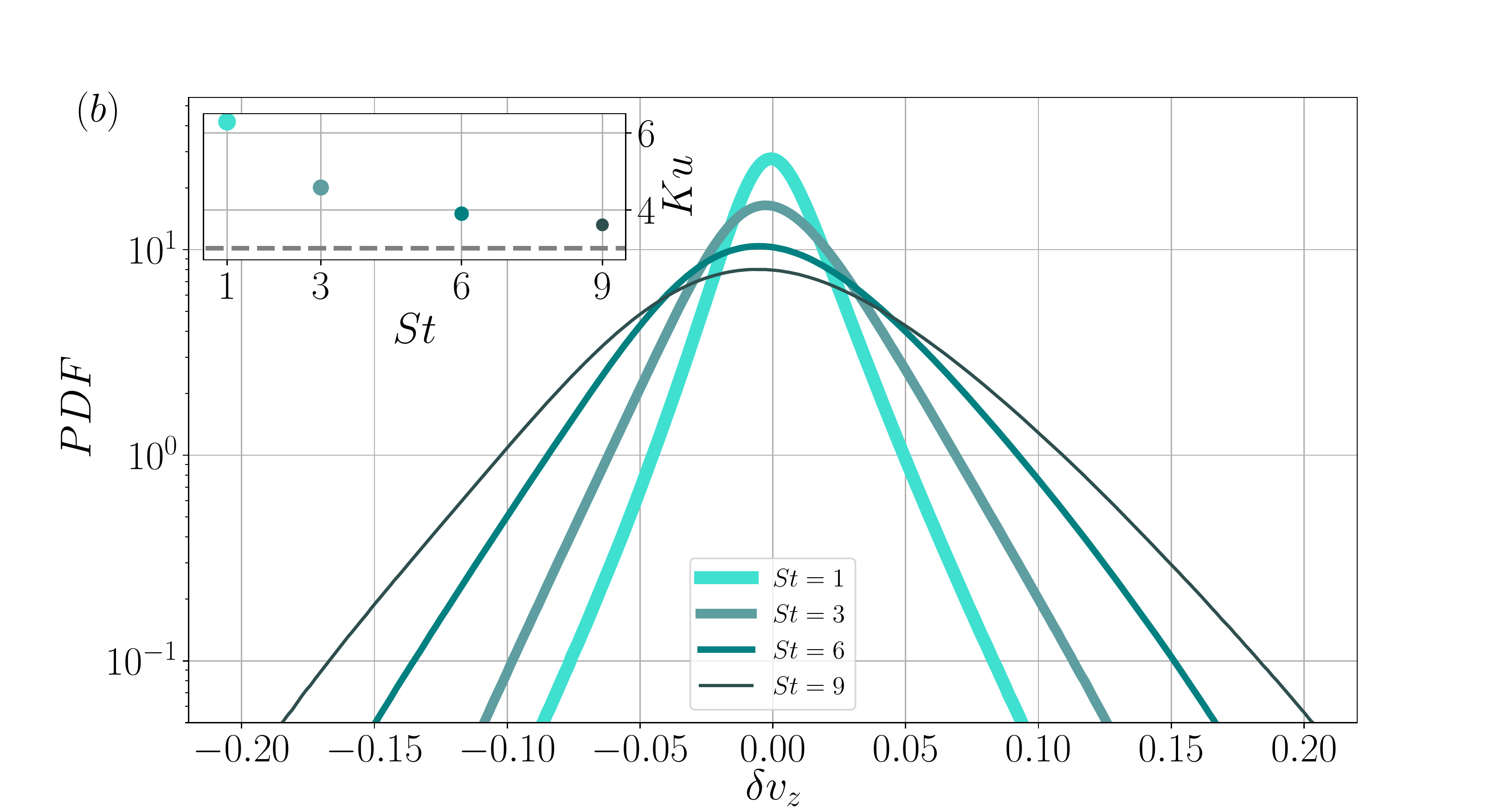}
\caption{{\it (a)} Particles' mean vertical velocity as a function of Fr for $\textrm{St}=6$. \AD{Error bars for $95\%$ confidence intervals are shown; standard deviations of fluctuations of $\left<v_z\right>$ in time are less than $2\%$.} The inset shows the anomaly $\left<v_z\right>/v^*$. {\it (b)} PDFs of particles' vertical velocity increments. As St increases, the PDFs are closer to Gaussian (see the kurtosis $Ku$ in the inset, $Ku=3$ is marked as a reference).}
\label{fig:dv}
\end{figure}

In Fig.~\ref{fig:dv}(a) we show the particles' mean vertical velocity $\left<v_z\right>$ for different Fr and fixed St; their values are compared to the Stokes terminal velocity $v^*$ in the inset. A discrepancy of about $25\%$ is observed for $\textrm{Fr} = 1.36$ (see inset), and particles in average tend to fall faster than $v^*$ except for cases with low Fr which can fall up to $5\%$ slower than $v^*$ (cumulants were studied to ensure the statistical significance of these values). Velocity fluctuations of the particles are also affected by St and Fr. In Fig.~\ref{fig:dv}(b) we show the probability density functions (PDFs) of velocity increments $\delta v_z = v_z(t+\tau/8) - v_z(t)$ (also a proxy of particles' vertical acceleration) for $\textrm{Fr} = 1.36$. As St increases the PDFs become more Gaussian, as also observed in DNSs of inertial particles without gravity \cite{bib:bec2006_JFM}. This also results from heavier particles being less affected by fluctuations in the flow velocity \cite{Bec2014}.

\begin{figure}
\includegraphics[width=7.8cm,trim={0 .8cm 0 0},clip]{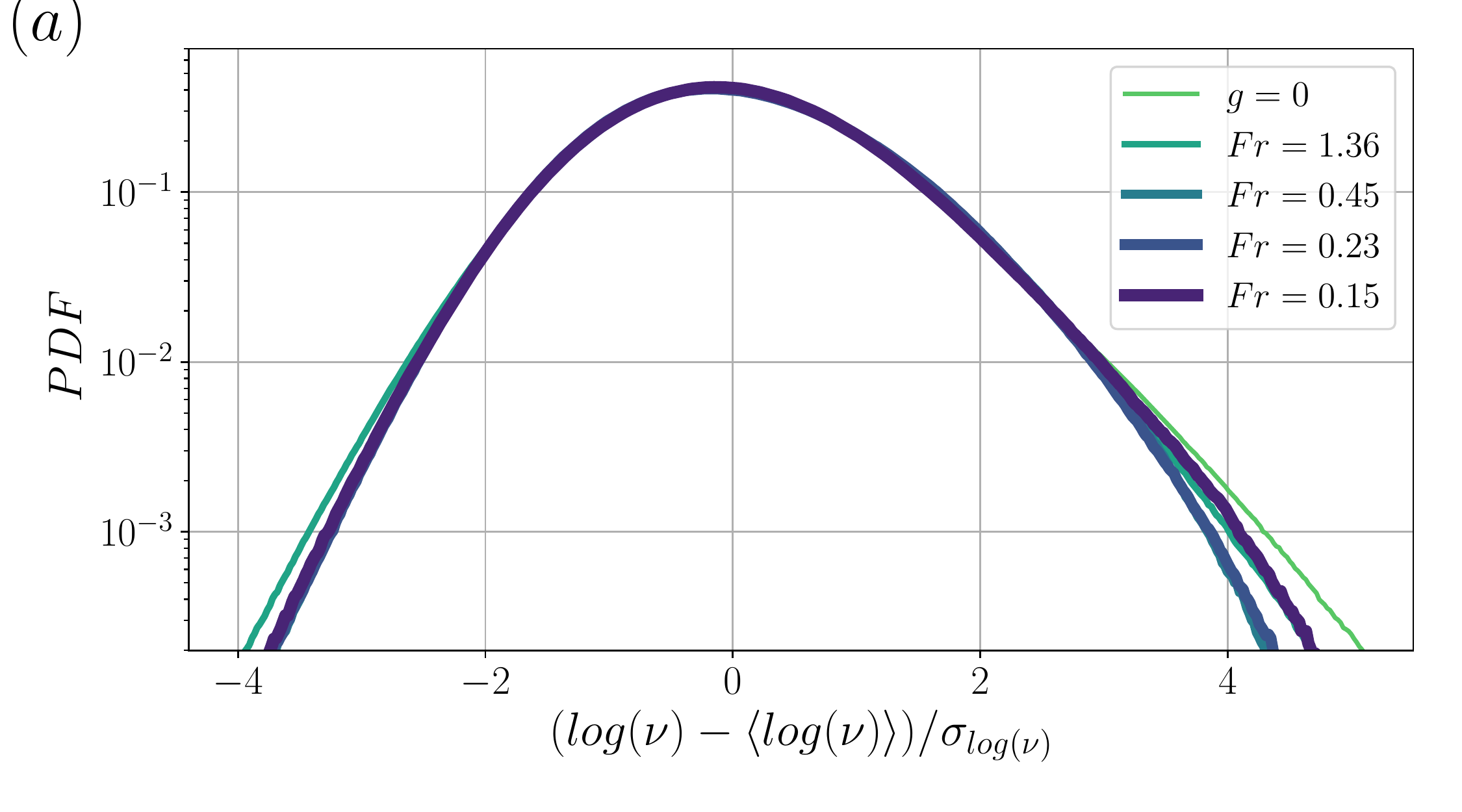}
\includegraphics[width=8.8cm,trim={0 1.2cm 0 .2cm},clip]{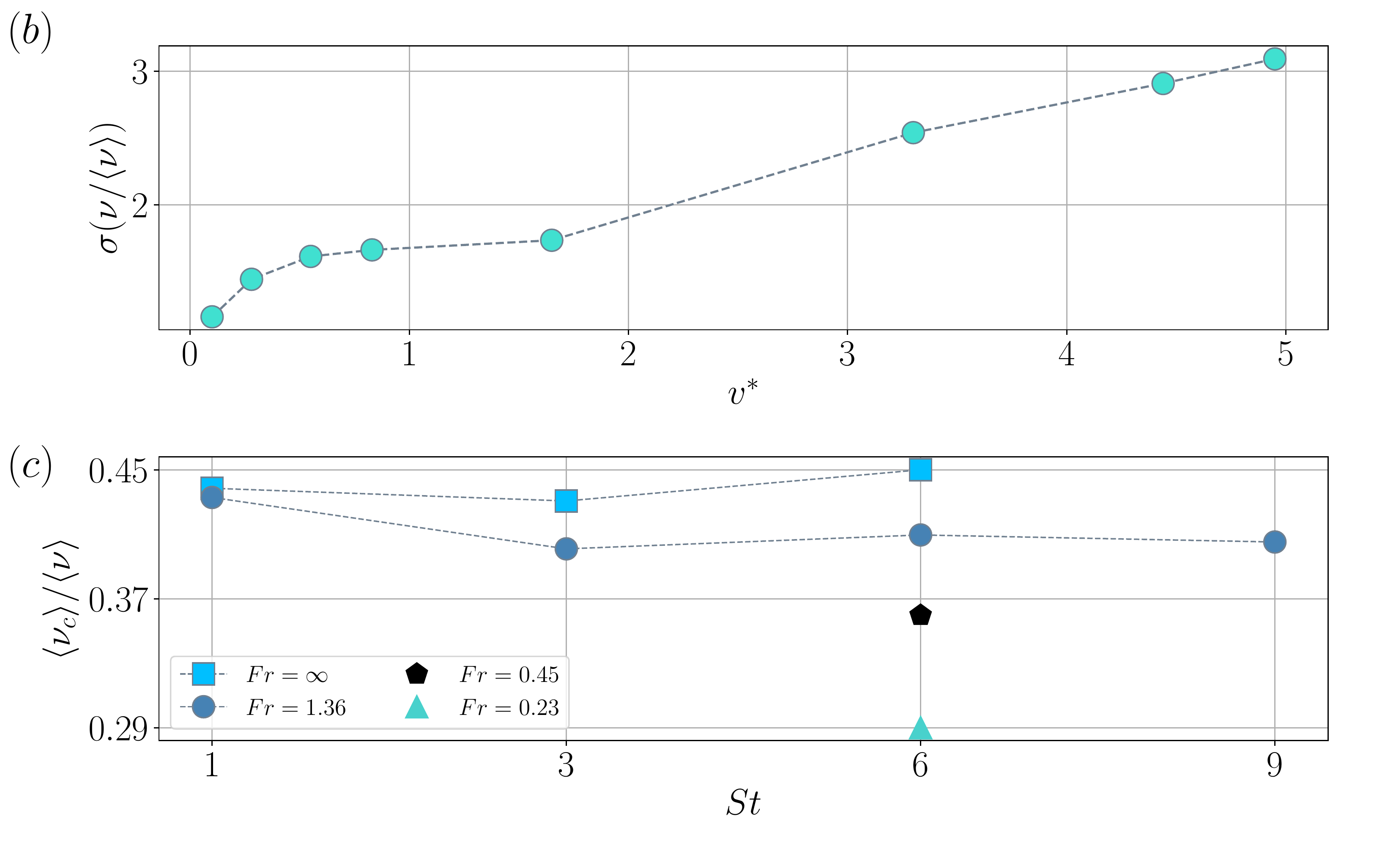}
\caption{{\it (a)} PDFs of Vorono\"i volumes for particles with $\textrm{St}=6$ and different Fr. {\it (b)} Standard deviation of the PDFs of $\nu/\langle\nu\rangle$ as a function of the theoretical terminal velocity $v^*$. {\it(c)} Mean  Vorono\"i volumes of clustered particles as a function of St and for different Fr.}
\label{fig:clusterpdf}
\end{figure}

The analysis so far does not consider that particles sample preferentially some flow regions. To quantify clustering we use three-dimensional Vorono\"i tessellation \cite{Tanemura}. Each cell has an associated particle, and the tessellation allows us to identify clusters with the cells with volume $\nu$ smaller than the mean of all volumes $\left<\nu\right>$. While other criteria can be used \cite{Monchaux2010, Obligado2014}, the one used here provides a simple way to discriminate between particles that are close together from other particles, as the volume of a Vorono\"i cell is inversely proportional to particle concentration. We can then cross-correlate clustered particles with physical magnitudes at the particles' positions.

Figure \ref{fig:clusterpdf}(a) shows the PDF of the Vorono\"i volumes for particles with $\textrm{St}=6$ and various Fr. In the absence of gravity a log-normal distribution was reported in experiments and simulations \cite{Obligado2014}; our results are in agreement with these studies. As seen in Fig.~\ref{fig:clusterpdf}(b), as gravity (or the Stokes terminal velocity) increases, the variance $\sigma
^2$ of the PDFs of $\nu/\langle\nu\rangle$ also increases, indicating particles accumulate in smaller clusters and preferential concentration becomes more dominant according to the criteria in \cite{Monchaux2010} (see also \cite{SM}). This is also confirmed by Fig.~\ref{fig:clusterpdf}(c), which shows that the average Vorono\"i volume of the clustered particles ($\nu_c$) is reduced as gravity increases (note this mean volume varies only slowly with St, but for fixed St decreases rapidly as Fr decreases). What regions of the flow do particles explore then? 

\begin{figure}
\begin{center}
\includegraphics[width=7cm,trim={0 0.95cm 0 .7cm},clip]{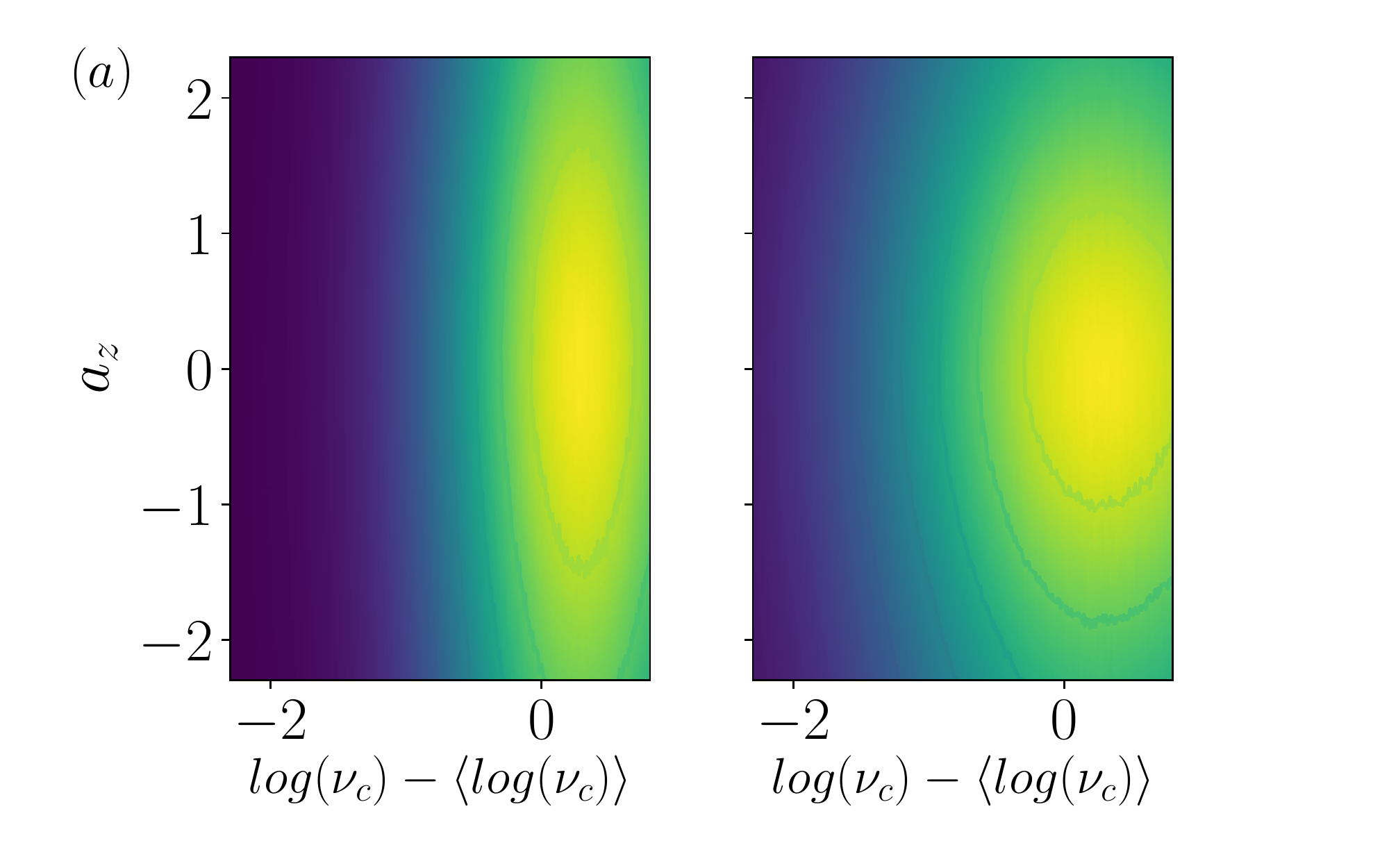}
\includegraphics[width=8.8cm,trim={0 1.4cm 0 0},clip]{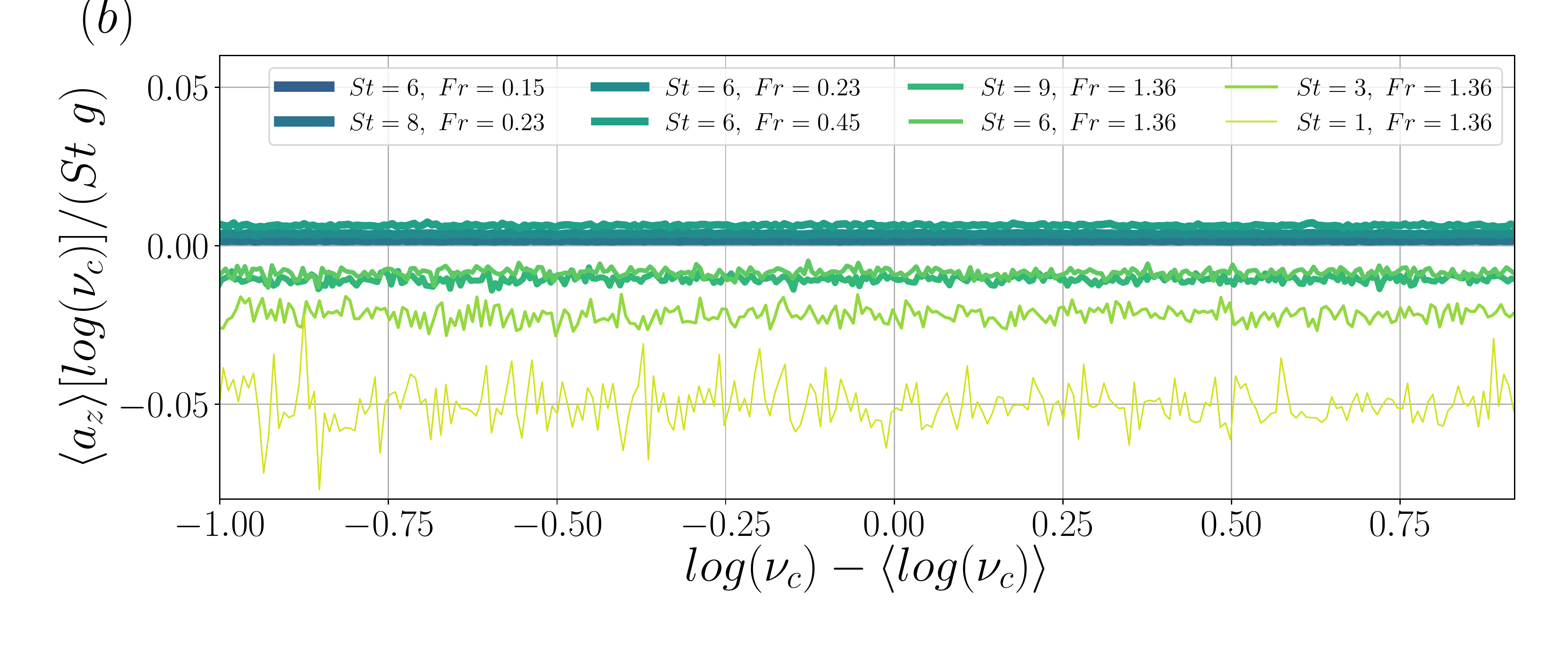}
\end{center}
\caption{{\it (a)} Joint PDFs of the volumes of cells belonging to clusters $\nu_c$, and of the vertical Lagrangian acceleration at the particles' positions, for $\textrm{St}=6$, and for two values of Fr (from left to right, $\infty$ and $0.23$). {\it (b)} Mean value of the vertical Lagrangian acceleration at the clustered particles' positions, as a function of $\log(\nu_c)$.}
\label{fig:jointst6}
\end{figure}

To answer this question we study at what points clusters are present. Having in mind classical centrifugation and sweep-stick mechanisms, this is done by computing cross-correlations between Vorono\"i volumes of clustered particles, and fluid Lagrangian acceleration and vorticity at Vorono\"i cells' center positions (i.e., at clustered particle's positions). As shown in \cite{SM}, no strong correlation is observed with the vorticity. Instead, in Fig.~\ref{fig:jointst6}(a) the joint PDFs of Vorono\"i volumes and of vertical Lagrangian accelerations are shown for cases with $\textrm{St}=6$, and $\textrm{Fr}=\infty$ (i.e., $g=0$) and $0.23$. In both cases clusters tend to be in regions of low fluid Lagrangian acceleration, and as was noted earlier, the clusters become denser with decreasing Fr (i.e., the maximum of the PDF moves to smaller values of $\nu_c$). However, there is a weak dependence of the mean Lagrangian acceleration of fluid elements in which particles accumulate when $g\neq 0$, as shown in Fig.~\ref{fig:jointst6}(b). While horizontal components of the fluid Lagrangian acceleration at clustered particles average to zero (not shown), the mean value of the vertical Lagrangian acceleration preferentially sampled by  particles with small St and large Fr tends to be negative, while particles with large St and small Fr prefer regions with small but positive $\left<a_z\right>$. This is independent of $\nu_c$, except for very small or very large clustered cells for which fluctuations appear as a result of small sampling sizes. 

\begin{figure}
\includegraphics[width=8.8cm,trim={0 1.3cm 0 0},clip]{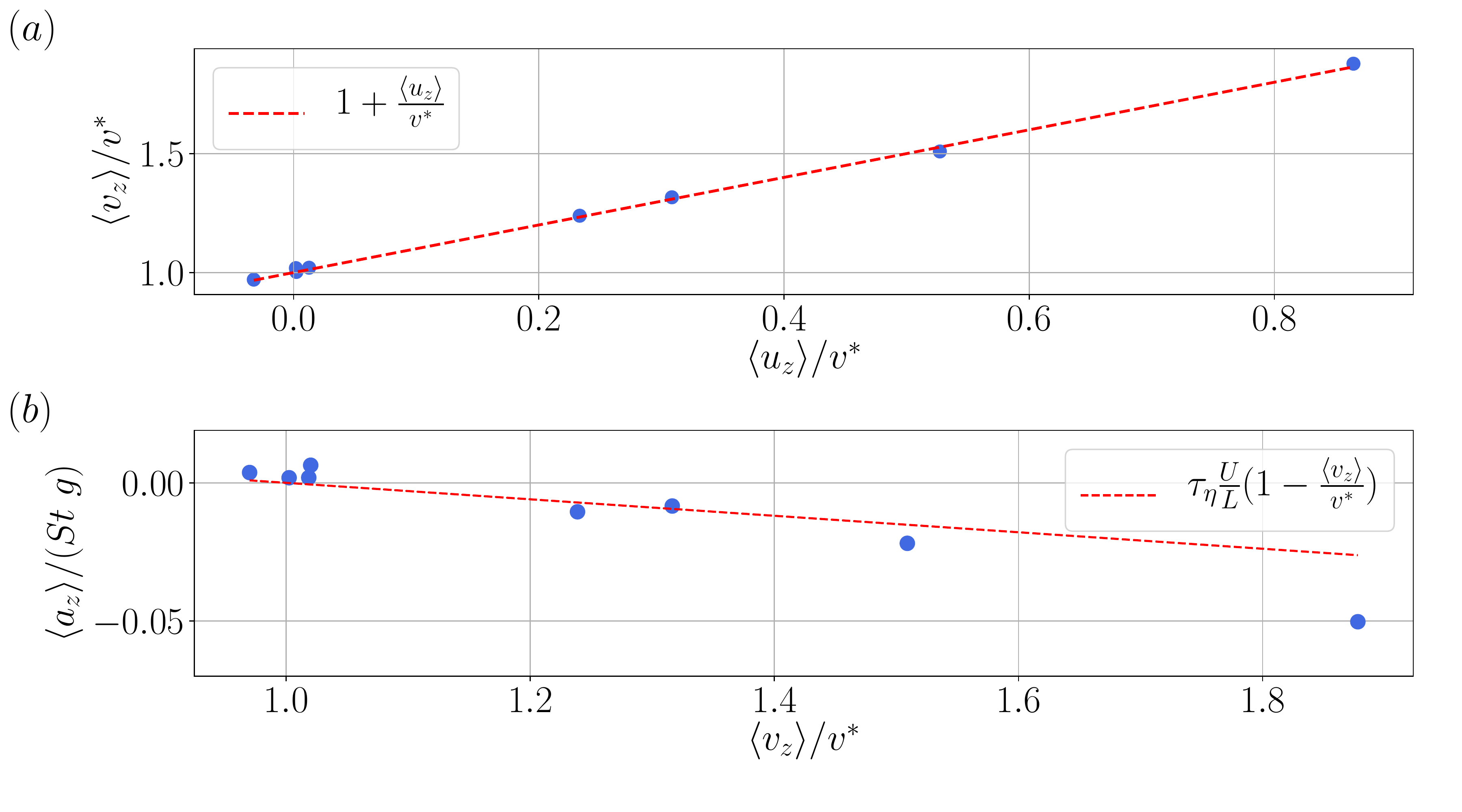}
\caption{{\it (a)} Mean vertical velocity of clustered particles as a function of the mean vertical velocity of fluid elements at particle's locations. Velocities are normalized by $v^*$; as a result points with $\left< v_z \right>/v^*>1$ (most of them)  indicate particles falling faster than the Stokes terminal velocity. The dashed line corresponds to Eq.~\eqref{eq:mean_vel}. {\it (b)} Same for the Lagrangian acceleration of fluid elements normalized by $\textrm{St} \, g$ as a function of  $\left< v_z \right>/v^*$. The dashed line corresponds to Eq.~\eqref{eq:mean_acc}.}
\label{fig:a(v)}
\end{figure}

Our results indicate that even in presence of gravity particles with $St>1$ preferentially sample low fluid acceleration regions, with a small positive or negative vertical bias depending on St and Fr, thus revalidating the sweep-stick mechanism initially proposed for situations without gravity. To explain the small but systematic deviations from accumulation in points with $\left<a_z\right>=0$ when $g \neq 0$, we now present a simple model that captures all cases studied. When particles reach a steady regime (i.e., a terminal velocity), they fulfil $\left<dv_z/dt\right>=0$. The averaged $z$-component of Eq.~\eqref{eq:inertial} then reduces to
\begin{equation}
  \langle v_z \rangle/v^* = 1 + \langle u_z\rangle/v^*,
  \label{eq:mean_vel}
\end{equation}
where the average is over particles. In other words, in the terminal regime, the average slippage velocity between the particles and the fluid is the Stokes terminal velocity $v^*$. As shown in Fig.~\ref{fig:a(v)}(a) for clustered particles, this relation is indeed satisfied by all cases studied.

Besides, one can relate the average Lagrangian acceleration of the carrier flow and its Lagrangian velocity by making the approximation that 
$\langle a_z\rangle \sim \langle u_z \rangle/\tau_L$, where $\tau_L=L/U$ is the eddy turnover time at the flow integral scale (note clusters and columns have global correlation length close to $L$, see Fig.~\ref{fig:columns} \AD{and \cite{SM}}). Introducing a Stokes number based on
$\tau_L$, $\textrm{St}_L=\tau/\tau_L$, we can estimate the term $\langle u_z\rangle/v^*$ in Eq.~\eqref{eq:mean_vel} using $\langle u_z\rangle /(g\tau) \sim \langle a_z \rangle/(g \, 
  \textrm{St}_L)$. As $\textrm{St}_L/\textrm{St} = \tau_\eta U/L$, then
\begin{equation}
  \langle a_z\rangle / (g \, St) = (\tau_\eta U/L) ( 1-\langle v_z\rangle/v^*).
  \label{eq:mean_acc}
\end{equation}

Figure \ref{fig:a(v)}(b) shows that our data is in reasonable agreement with Eq.~\eqref{eq:mean_acc} (with the exception of the case with $\textrm{St}=1$ for which centrifugation and sweep-stick mechanisms may be competing, see also \cite{SM}). Note that in both figures we have a loitering case with $\langle v_z\rangle/v^* < 1$. The scaling also explains observations in Figs.~\ref{fig:dv} and \ref{fig:jointst6}: Particles tend to stick to (or to preferentially fall through) fluid elements with Lagrangian acceleration which is small and dependent on the ratio  $\left< v_z \right>/v^*$, which from our data seems to also depend on St and Fr. In fact, Eq. ~\eqref{eq:mean_acc} can be rewritten as $\langle a_z\rangle/g = St_L(1- \langle v_z\rangle/v^*)$. The range of validity of this equation is then $St_L<1$ (as the average lifespan of a zero-acceleration point is typically of the order of $\tau_{L}$ \cite{Coleman2009}) and $St>1$ (for the ``stick'' mechanism of such points to be active). For $g=0$, Eq.~\eqref{eq:mean_acc} reduces to $\langle a_z\rangle=0$, as in the usual sweep-stick mechanism. Finally, note the scaling in Eq.~\eqref{eq:mean_acc} is stationary: particles are not accelerated and reach a terminal velocity $\langle v_z\rangle$ even though they preferentially cross regions of the carrier flow that may be slightly accelerated.

The results indicate that small and heavy inertial particles in a turbulent flow with linear drag and gravity accumulate in isotropic clusters (when $g$ is sufficiently small), or fall creating columns which result in regions of even stronger particle accumulation. In the latter case there is a pronounced anisotropy in the distribution of particles in the direction of gravity. The time that takes particles to reach their terminal velocity (controlled by the Stokes time and Froude numbers) affects the formation and width of these columns. Moreover, the particles' terminal velocity presents an anomaly with respect to the Stokes terminal velocity in a viscous fluid at rest. The study of the joint  statistics of clustered particles and of fluid Lagrangian acceleration at points of aggregation supports a modified sweep-stick mechanism in which particles tend to concentrate in points of low Lagrangian acceleration, where they fall with a velocity such that the relative mean slippage velocity becomes equal to the Stokes terminal velocity. This anomaly in the terminal velocity results in the small deviation from accumulation in zero-acceleration points of the carrier fluid. In the limit of $g=0$ the Stokes terminal velocity (and hence the slippage) vanishes, and classical sweep-stick is recovered, in which particles are stuck to actual zero-acceleration points. Such a generalization of the sweep-stick argument to heavy particles with gravity is in good agreement with recent experimental data \cite{bib:sumbekova2017_PRF}, and opens the door to study other cases in which particles transported by fluids are subjected to external acceleration forces as, e.g., in active matter \cite{durham13, gustavsson16}.

\begin{acknowledgments}
{\it This project was supported by the IDEXLyon project (contract ANR-16-IDEX-0005) under the auspices of University of Lyon. This work has also been partially supported by the ECOS project A18ST04. PDM acknowledges support from grant PICT No.~2015-3530.}
\end{acknowledgments}

\bibliography{ms}

\clearpage

\onecolumngrid
\section*{Supplemental Material: Preferential concentration of free-falling heavy particles in turbulence}

\section{Characteristic scales}

In this work we consider direct numerical simulations of point particles carried by an incompressible turbulent flow in a $(2\pi)^3$ cubic periodic domain. Details on the numerical method used to solve the incompressible Navier-Stokes equations and the equations for the inertial particles can be found in \cite{mininni2011hybrid}. To define Reynolds numbers we use a characteristic one-component velocity $u'=U/\sqrt{3}$, where $U = (\left<|{\bf u}|^2\right>)^{1/2}= (2E)^{1/2}$ is a r.m.s.~velocity computed from the Eulerian fluid velocity in the simulations, and where $E$ is the fluid mean kinetic energy. The flow integral and Taylor scales are then defined respectively as
\begin{equation}
    L = \frac{2 \pi}{E} \int{\frac{E(k)}{k} \, dk} , \,\,\,\,\,\,\,\,\,\,\,\,\,\,
    \lambda = \left( \frac{15 \nu {u'}^2}{\varepsilon} \right)^{1/2},
\end{equation}
where $E(k)$ is the isotropic energy spectrum in terms of the wavenumber $k$, $\varepsilon$ is the energy dissipation rate, and $\nu$ the fluid kinematic viscosity. The Reynolds number is then given by $\textrm{Re}=u'L/\nu \approx 5700$, while the Taylor-scale Reynolds number is $\textrm{Re}_\lambda=u'\lambda/\nu\approx 300$.

\section{Statistics of carrier fluid Lagrangian acceleration and vorticity}

In Fig.~4(a) of the letter we showed the correlation between clustered particles and vertical Lagrangian acceleration of fluid elements at particles' positions. Here we provide supplemental information of the correlation of particles' positions with the carrier fluid Lagrangian acceleration and vorticity, to separate between sweep-stick and centrifugal expulsion mechanisms. 

\begin{figure}[b]
\includegraphics[width=14.8cm,trim={.8cm 1.2cm 0 1.2cm},clip]{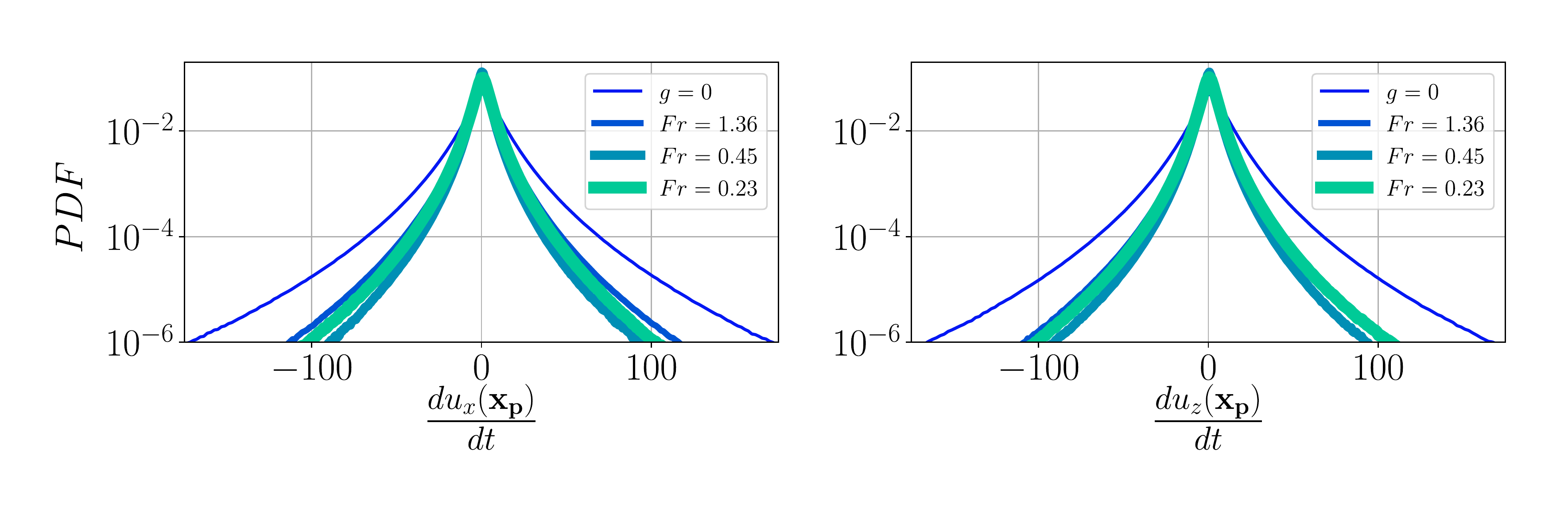}
\caption{PDFs of the carrier fluid Lagrangian acceleration ($x$ and $z$ components) at all particle's positions, for inertial particles with $\textrm{St}=6$.}     
\label{fig:lacc_st6}
\end{figure}

The probability density functions (PDFs) of the Lagrangian acceleration of the carrier fluid elements at all particles' positions is centered around zero for all cases studied, indicating that there are more particles in points of low Lagrangian acceleration (see Fig.~\ref{fig:lacc_st6}, for different Froude numbers and at a fixed Stokes number of $\textrm{St}=6$, and where $d/dt$ is the convective derivative). Moreover, it is also observed (but not shown here) that at fixed Fr and as St increases, these PDFs tend to become narrower, indicating particles explore less regions with extreme Lagrangian accelerations.

\begin{figure}
\includegraphics[width=14.8cm,clip]{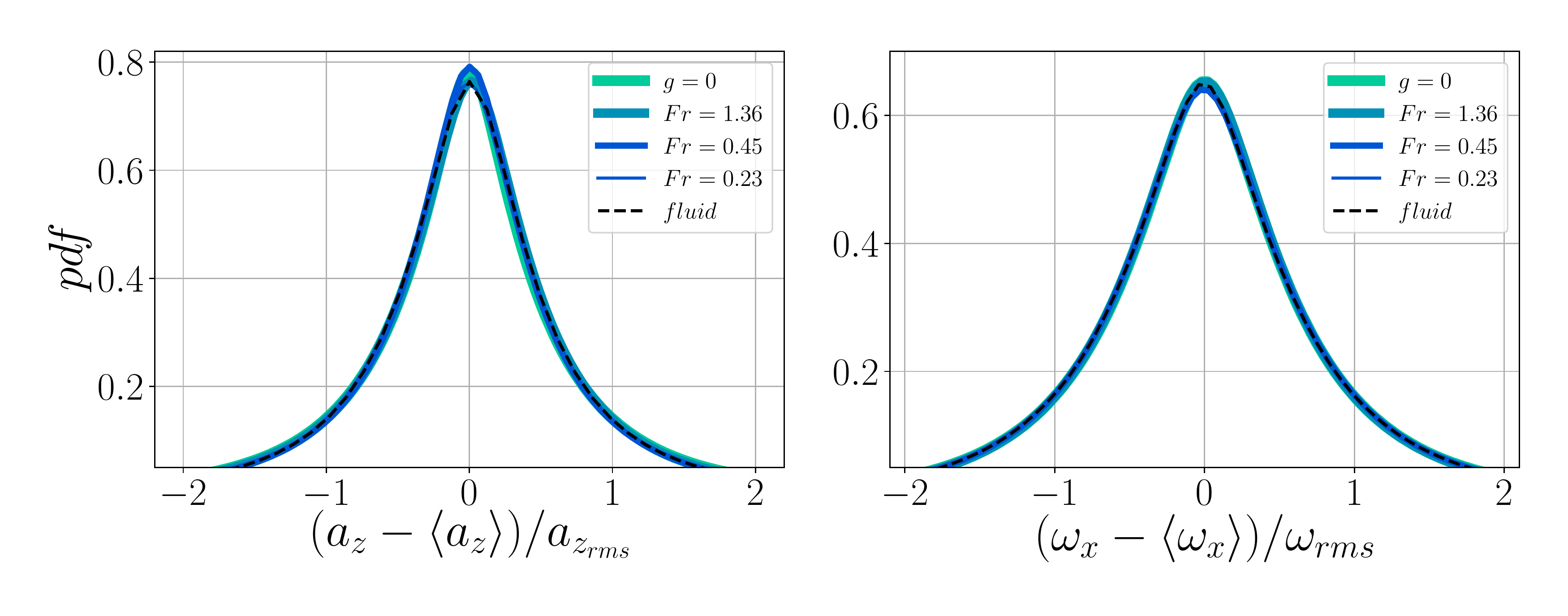}
\caption{{\it Left:} PDFs of the fluid vertical Lagrangian acceleration  at all inertial particle's positions. {\it Right:} PDFs of the $x$ component of the fluid vorticity at all inertial particle's positions. The dashed lines show the PDFs for fluid elements at tracers' positions. All quantities have their mean subtracted, and are normalized by their r.m.s.~values.}
\label{fig:sm_fig1}
\end{figure}

To study in more detail the sweep-stick mechanism, we compare normalized PDFs of Lagrangian acceleration and of vorticity at the positions of all inertial particles, against those of tracer fluid elements, as done, e.g., in the case without gravity in \cite{Coleman2009}. Figure \ref{fig:sm_fig1} shows these PDFs (with the mean value subtracted to correct for modified sweep-stick mechanisms when $g\neq 0$, and normalized by their r.m.s.~values) for the vertical component of the Lagrangian acceleration ($a_z=du_z({\bf x}_p)/dt$) and for a horizontal component of the vorticity ($\omega_x$). Similar results are obtained for other components of these quantities. Again, we consider different values of Fr for a fixed Stokes number $\textrm{St}=6$, and we also compare with the PDFs of these quantities at the position of $10^6$ tracers. To this end we injected Lagrangian tracers in the fluid (that do not accumulate preferentially), and integrated them following the equation $d\mathbf{x}_\textrm{tr}/dt=\mathbf{v}_\textrm{tr}(t)=\mathbf{u}(\mathbf{x}_\textrm{tr},t)$, where $\mathbf{v}_\textrm{tr}$ is the tracer's velocity, $\mathbf{x}_\textrm{tr}$ its position, and $\mathbf{u}(\mathbf{x}_\textrm{tr},t)$ is the fluid velocity.

\vspace{5pt}
As shown in previous studies \cite{Coleman2009}, the peaks of the PDFs of Lagrangian acceleration of fluid elements conditioned to the inertial particles' positions are larger and narrower than that of the tracers. In particular, for $g=0$ we recover previous results. When different values of Fr are considered, an ``imperfect" sweep-stick can be observed, as there is a larger concentration of particles in points with $a_z=\left<a_z\right>$. We also show in Fig.~\ref{fig:sm_fig1} the PDFs of the $x$ component of the vorticity of the fluid in all inertial particles' positions for the aforementioned cases. All the curves collapse into one regardless of the Froude number, have a broader peak, and also collapse with the corresponding PDF of the tracers. Thus, we conclude there is no strong correlation of the position of inertial particles with the vorticity.

\section{Supplemental information on Vorono\"i volumes}
\begin{figure}[h]
\includegraphics[width=10.8cm,clip]{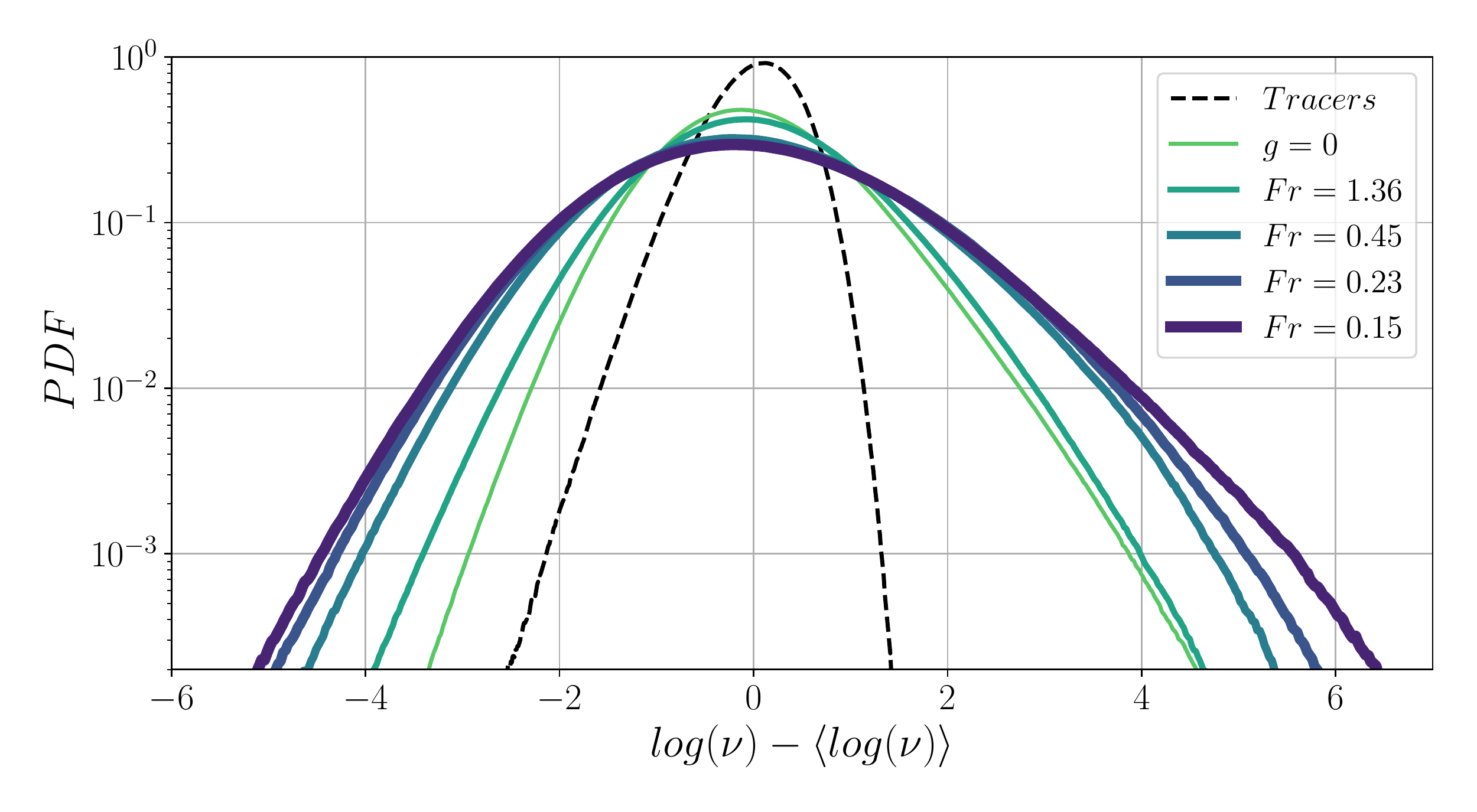}
\caption{PDFs of the logarithm of Vorono\"i volumes $\nu$ for different values of Fr at a fixed Stokes number $\textrm{St}=6$, not normalized by their dispersion, and compared with those of tracers.}
\label{fig:sm_fig2}
\end{figure}
\vspace{3pt}

In figure \ref{fig:sm_fig2} we show the PDFs of the Vorono\"i volumes $\nu$ for different values of Fr at a fixed Stokes number $\textrm{St}=6$, compared to the Vorono\"i volumes obtained for Lagrangian tracers. Compared with Fig.~3(a) in the letter, PDFs in this figure are not normalized by their dispersion to see the effect of changing Fr. Previous studies have also shown that the Vorono\"i volumes of tracers  follow a Random Poisson Process (RPP) distribution, with a standard deviation of $\sigma = 0.423$ for the three-dimensional case, a value that we recover (see, e.g., \cite{Tanemura}).

Note that as the Froude number decreases, the PDFs of inertial particles display a larger dispersion, and when compared with the RPP distribution, this results in a larger number of concentrated particles (i.e., of probability of finding particles with volumes smaller than that predicted by the RPP). As the Froude number decreases there are also more volumes with extreme values (compared with the tracers), i.e., there are more voids as well as more clusters.

\section{Regression analysis for the modified sweep-stick mechanism}

\begin{figure}[t]
\includegraphics[width=10.8cm,clip]{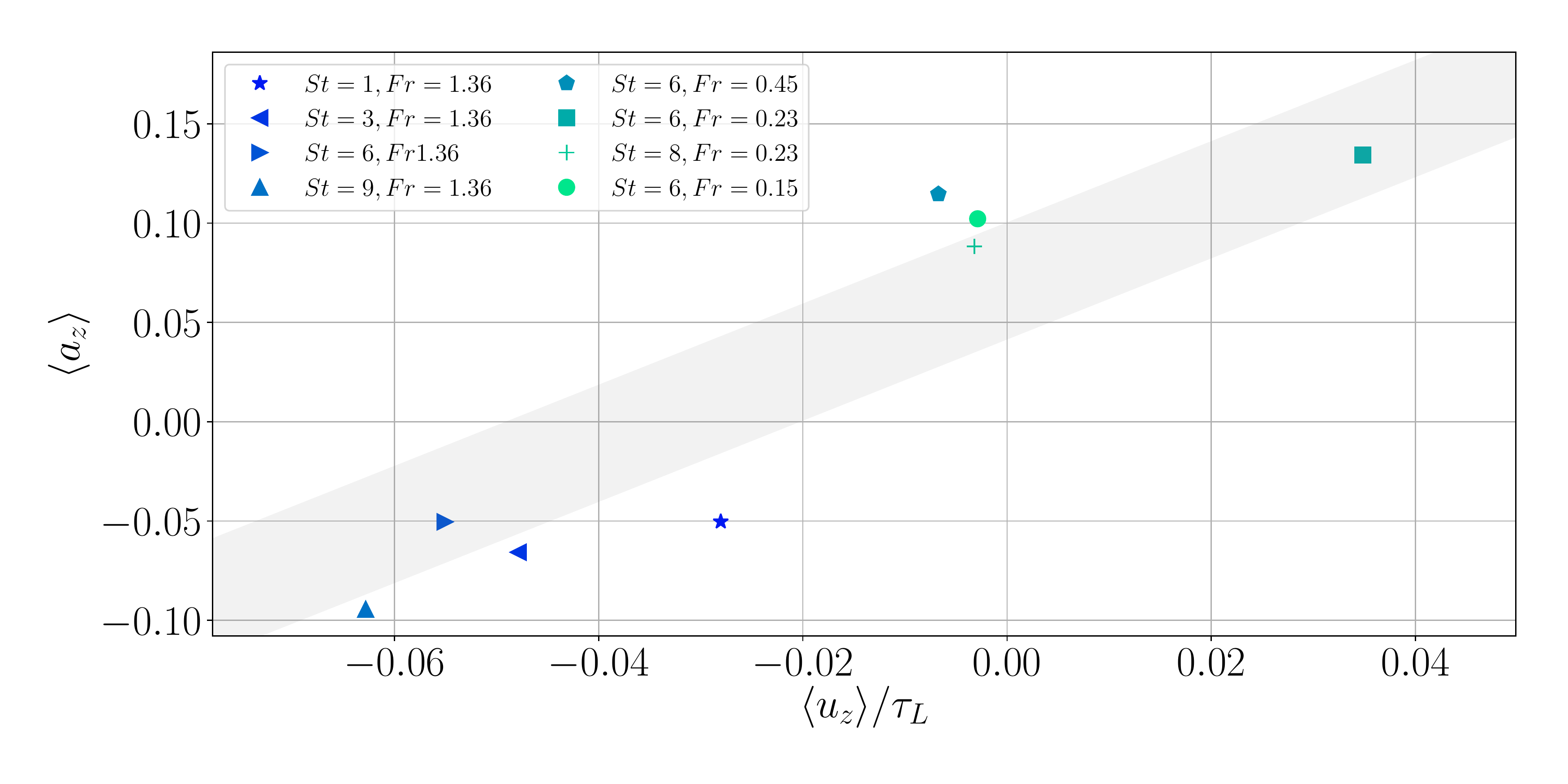}
\caption{Mean vertical component of the Lagrangian acceleration of the fluid vs.~the mean vertical fluid velocity divided by the integral-scale eddy turnover time, at the clustered particles' positions, and for different values of St and Fr. The dashed thick line indicates a slope of one.}
\label{fig:sm_fig3}
\end{figure}

The modified sweep-stick mechanism assumes that $\langle a_z\rangle \sim \langle u_z \rangle/\tau_L$, where $\tau_L$ is the eddy turnover time at the flow integral scale. While this argument is based on the observation that  columns have global correlation length close to $L$, this correlation can be also studied directly in the numerical data. In Fig.~\ref{fig:sm_fig3} we show the mean vertical Lagrangian acceleration of the fluid in the clustered particles' positions, $\left<a_z\right>$, as a function of the mean fluid vertical velocity at the same positions, $\left<u_z\right>$, divided by the eddy turnover time $\tau_L$. These two quantities display a reasonable correlation with a slope of $\mathcal{O}(1)$. Moreover, the figure also indicates a relevant parameter in the preferential selection by the particles of regions of the flow with positive or negative mean vertical velocity (or acceleration): Points at the left of the figure have larger Fr, while points at the right have smaller Fr.

To verify the goodness of fit of the relation given by Eq.~(3) in the letter,
$$\frac{\langle a_z\rangle}{g \, St} = \frac{\tau_\eta U}{L}\left(1- \frac{\langle v_z\rangle}{v^*}\right),$$
we also computed regression analysis on $\langle a_z\rangle/(g \, St)$ and $\langle v_z\rangle/v^*$. Firstly, the best correlation and collapse of data was obtained when $\langle a_z\rangle$ was normalized by $g \, St$ (see Fig.~5(b) in the letter). Secondly, we obtained a Pearson's correlation coefficient between $\langle a_z\rangle/(g \, St)$ and $\langle v_z\rangle/v^*$ (in absolute value) of $0.98$, with a $p$-value for the linear relation of $1.02 \times 10^{-5}$. As a reference, assuming a cuadratic relation between these two quantities (or equivalently, between the two quantities in Fig.~5(b) of the letter) yields a Pearson's correlation coefficient of $0.88$ and a $p$-value of $3.61 \times 10^{-3}$.

\end{document}